\documentclass[%
reprint,
showkeys,
nofootinbib,
amsmath,amssymb,
aps,
prl,
]{revtex4-1}
\usepackage[marginal]{footmisc}
\usepackage{graphicx}
\usepackage{dcolumn}
\usepackage{bm}
\usepackage[colorlinks,allcolors=black,CJKbookmarks=True]{hyperref}

\begin{document}
	
	\title{Distinguishing the nanohertz gravitational-wave sources by the observations of compact dark matter subhalos}
	
	\author{Jing Liu$^{1,2}$}
	\email{liujing@ucas.ac.cn}
	
	\affiliation{$^{1}$International Centre for Theoretical Physics Asia-Pacific, University of Chinese Academy of Sciences, 100190 Beijing, China}
	
	\affiliation{$^{2}$Taiji Laboratory for Gravitational Wave Universe~(Beijing/Hangzhou), University of Chinese Academy of Sciences, 100049 Beijing, China}
	
	\begin{abstract}
		The latest pulsar timing array data reveals evidence of nanohertz gravitational waves~(GWs), which have been explained by both cosmological and astrophysical sources. However, current observations lack the precision needed to differentiate between different models from the spectral index. We find that the cosmological GW sources, including bubble collisions, sound waves, domain walls, condensate fragmentations, and primordial curvature perturbations, induce large energy density perturbations so that most dark matter will exist in gravitationally self-bound subhalos. 
		Then, the observation of such substructures of dark matter can serve as a novel independent method to confirm or exclude the cosmological GW sources.
	\end{abstract}
	
	\maketitle
	\emph{Introduction}. 
	The direct observation of gravitational waves (GWs) provides a novel insight into understanding the early history of the Universe and discovering new fundamental physics beyond the Standard Model~(SM)~\cite{Maggiore:1999vm,Boyle:2005se,Caprini:2018mtu,Cai:2017cbj,Bian:2021ini}. A stochastic gravitational-wave background (SGWB) is formed from the superposition of GWs originating from numerous uncorrelated sources. Multiband GW observations from LIGO/Virgo/KAGRA~\cite{TheLIGOScientific:2014jea,VIRGO:2014yos,Somiya:2011np}, LISA~\cite{Audley:2017drz}, Taiji~\cite{Guo:2018npi}, TianQin~\cite{TianQin:2015yph}, IPTA~\cite{Hobbs:2009yy} and SKA~\cite{Carilli:2004nx} provide an extensive approach to explore new physics over a wide range of energy scales. The evidence of nanohertz GWs is collectively comfirmed by CPTA~\cite{Xu:2023wog}, NANOGrav~\cite{NANOGrav:2023hvm}, PPTA~\cite{Reardon:2023gzh} and EPTA~\cite{Antoniadis:2023ott}, which has been interpreted as SGWBs from different sources in the early Universe~\cite{NANOGrav:2023hvm,Ellis:2020ena,Blasi:2020mfx,DeLuca:2020agl,Cai:2020ovp,Buchmuller:2020lbh,Addazi:2020zcj,Ratzinger:2020koh,Nakai:2020oit,Samanta:2020cdk,Vagnozzi:2020gtf,Bian:2020urb,Neronov:2020qrl,Li:2020cjj,Kohri:2020qqd,Liu:2020mru,Zhou:2020kkf,Domenech:2020ers,Inomata:2020xad,Kitajima:2020rpm,Ramberg:2020oct,Pandey:2020gjy,Gao:2020tsa,Li:2021qer,Kawasaki:2021ycf,Blanco-Pillado:2021ygr,Sharma:2021rot,Sakharov:2021dim,NANOGrav:2021flc,Borah:2021ocu,Yi:2021lxc,Liu:2021svg,Cai:2021wzd,Gao:2021lno,Benetti:2021uea,Ashoorioon:2022raz,RoperPol:2022iel}. Then, determining the source of the SGWB is the next major objective. 
	However, the current observational precision is insufficient for model discrimination~\cite{Bian:2020urb}. In this $letter$, we propose a novel method that utilizes the observations of dark matter~(DM) subhalos to distinguish the nanohertz GW sources.
	
	
	The violent processes that efficiently produce GWs also in general induce large energy density perturbations. This has been demonstrated by various lattice simulations of cosmological GW sources, such as the collisions of bubble walls during the first-order phase transitions~(FOPTs)~\cite{Hindmarsh:2015qta,Cutting:2018tjt,Di:2020kbw}, the formation of topological defects~\cite{Hiramatsu:2013qaa,Press:1989yh,Allen:1990tv,Ringeval:2005kr}, and nonlinear evolution during the preheating process~\cite{Amin:2010dc,Zhou:2013tsa,Lozanov:2019ylm,Liu:2017hua,Fu:2019qqe}.
	These induced fluctuations may surpass those originating from inflation and become dominant at smaller scales, resulting in Cosmic Microwave Background~(CMB) spectral distortions~\cite{Chluba:2011hw,Chluba:2012we,Franciolini:2023lgy}, changes in primordial helium abundance~\cite{Jeong:2014gna,Nakama:2014vla,Inomata:2016uip} and the formation of DM subhalos~\cite{Bringmann:2011ut,Clark:2015tha,Ramani:2020hdo,Lee:2020wfn}. Particularly, we find that the observations of DM subhalos have the potential to constrain the GW sources in a wide range of energy scales up to the electroweak scale and give upper bounds on the SGWBs in the nHz and $\mu$Hz bands. We find that for various cosmological GW sources that are strong enough to explain the nanohertz SGWB, the corresponding halo abundance is close to unity. This is a distinct feature to be verified or excluded by various astrophysical observations~\cite{Munoz:2019hjh,Buschmann:2017ams,DiazRivero:2017xkd,VanTilburg:2018ykj,Mondino:2020rkn,Mishra-Sharma:2020ynk,Li:2012qha,Clark:2015sha,Dror:2019twh,Ramani:2020hdo,Siegel:2007fz,Baghram:2011is,Brandt:2016aco}, thus providing a novel method to distinguish between different GW sources. 
	For convenience, we choose $c=8\pi G=1$ throughout this $letter$.
	
	
	\emph{GWs versus energy density perturbations}.
	We present induced energy density perturbations of the cosmological GW sources as a comparison of their energy spectrum of GWs. We define the energy scale ($T_{*}$), the Hubble parameter ($H_{*}$), and the scale factor ($a_{*}$) at the GW production time for each source. Additionally, we assume that the effective relativistic degree of freedom at the production time ($t_{*}$) is $g_{*}=10$, while at present ($t_{0}$), $g_{0}=3.36$. The current radiation density fraction $\Omega_{\mathrm{r}}h^{2}$ is estimated as $4.2\times 10^{-5}$, the current Hubble constant is $H_{0}=67.3\,\mathrm{km/s/Mpc}$ and $h=0.673$.
	
	During the FOPTs, true vacuum bubbles nucleate via the quantum tunneling effect, and the vacuum energy transfers to bubble walls and the background radiation. The GW sources during the FOPTs mainly include bubble collisions, sound waves, and turbulence. Here, we do not consider GWs from turbulence because of the large theoretical uncertainty~(see Ref.~\cite{RoperPol:2019wvy,Brandenburg:2021bvg} for recent progress). We apply the fitting results of the GW energy spectrums from bubble collisions and sound waves given in Ref.~\cite{Huber:2008hg,Hindmarsh:2015qta,Guo:2020grp,Tamanini:2016zlh,NANOGrav:2021flc}, 
	\begin{equation}
		\begin{split}
			&\Omega_{\mathrm{GW}}^{\mathrm{bc}}h^{2}(f)=7.7\times 10^{-5}g_{*}^{-\frac{1}{3}}\left(\frac{\kappa_{\phi}\alpha_{*}}{1+\alpha_{*}}\right)^{2}\left(\frac{H_{*}}{\beta}\right)^{2}\\
			&\frac{0.48v_{w}^{3}}{1+5.3v_{w}^{2}+5v_{w}^{4}}\times 8\left[\left(f/f_{p}^{\mathrm{bc}}\right)^{-2}+3\left(f/f_{p}^{\mathrm{bc}}\right)^{\frac{2}{3}}\right]^{\frac{3}{2}}\,,
		\end{split}
	\end{equation}
	\begin{equation}
		\begin{split}
			&\Omega_{\mathrm{GW}}^{\mathrm{sw}}h^{2}(f)=7.7\times 10^{-5}g_{*}^{-\frac{1}{3}}\left(\frac{\kappa_{\mathrm{sw}}\alpha_{*}}{1+\alpha_{*}}\right)^{2}\left(\frac{H_{*}}{\beta}\right)0.513v_{w}\\
			&\left(1-(1+2\tau_{\mathrm{sw}}H_{*})^{-\frac{1}{2}}\right)\left(f/f_{p}^{\mathrm{sw}}\right)^{3}\left[7/\left(4+3\left(f/f_{p}^{\mathrm{sw}}\right)^{2}\right)\right]^{\frac{7}{2}}\,,
		\end{split}
	\end{equation}
	where $f$ is the frequency of GWs, $\alpha_{*}$ is the ratio of energy densities of vacuum and background radiation, $\beta$ is the inverse time of the FOPT duration, $v_{w}$ is the bubble wall velocity, $\kappa_{\phi}$ and $\kappa_{\mathrm{sw}}$ are the fractions of the vacuum energy that transform into bubble walls and sound waves. The peak frequencies are $f_{p}^{\mathrm{bc}}=1.13\times 10^{-10}\,\mathrm{Hz}\,\frac{0.35}{1+0.07v_{w}+0.69v_{w}^{4}}(\frac{\beta}{H_{*}})(\frac{T_{*}}{\mathrm{MeV}})(\frac{g_{*}}{10})^{1/6}$, and $f_{p}^{\mathrm{sw}}=1.13\times 10^{-10}\,\mathrm{Hz}\,\frac{0.536}{v_{w}}(\frac{\beta}{H_{*}})(\frac{T_{*}}{\mathrm{MeV}})(\frac{g_{*}}{10})^{1/6}$ for bubble collisions and sound waves. $\tau_{\mathrm{sw}}$ is the sound-wave lifetime.
	Recently, we propose that the FOPTs induce energy density perturbations due to the randomness of the quantum tunneling process. We utilize the semi-analytical result of $\delta_{H}$ from Ref.~\cite{Liu:2022lvz},
	where $\delta_{H}\propto \alpha_{*}\beta^{-5/2}$ under the condition $\alpha_{*}< 1$ and $\beta/H_{*}\gg 1$. 
	In the FOPT case, we restrict the parameters $v_{w}\sim 1$ and $\alpha_{*}<1$.
	
	Domain walls form in the early Universe as a consequence of the spontaneous breaking of discrete symmetries~\cite{Vilenkin:1984ib,Vilenkin:1981zs}. One can avoid the cosmic domain wall problem by introducing a bias term in the effective potential so that domain walls annihilate before dominating the Universe~\cite{Gelmini:1988sf,Larsson:1996sp}. 
	In Refs.~\cite{Hiramatsu:2013qaa,Saikawa:2017hiv}, the GW production from domain walls has been simulated in lattice, and the peak value of $\Omega_{\mathrm{GW}}$ has been determined as
	\begin{equation}
		\Omega_{\mathrm{GW},p}^{\mathrm{dw}}h^{2}=\Omega_{\mathrm{r}}h^{2}\left(\frac{g_{0}}{g_{*}}\right)^{1/3}\frac{ \tilde{\epsilon}_{\mathrm{GW}} \mathcal{A}^{2} \sigma^{2}}{24\pi H_{*}^{2}}\,,
	\end{equation}
	where $\sigma$ is the domain wall tension, $\mathcal{A}\approx 0.8$ and $\tilde{\epsilon}_{\mathrm{gw}} \simeq 0.7$ are constants fixed by simulations. The peak frequency is $f^{\mathrm{dw}}_{p}=\left(\frac{H_{*}^{2}}{H_{0}^{2}\Omega_{\mathrm{r}}}(\frac{g_{*}}{g_{0}})^{1/3}\right)^{-1/4}H_{*}$. $\Omega_{\mathrm{GW}}^{\mathrm{dw}}(f)$ is proportional to $f^{3}$ and $f^{-1}$ for $f\ll f^{\mathrm{dw}}_{p}$ and $f\gg f^{\mathrm{dw}}_{p}$.
	The irregular distribution of domain walls also induces density perturbations at around the Hubble horizon scale and the amplitude of $\delta_{H}$ is close to the domain wall energy density, $\delta_{H}\approx \mathcal{A}\sigma/t_{*}$~\cite{Press:1989yh,Saikawa:2017hiv}.
	
	
	For local strings without long-range interactions, the production of massive excitations from long wavelength oscillation modes is suppressed~\cite{Martins:2003vd,Olum:1999sg} so that the GW production efficiency is very high. Therefore, string loops can produce strong GW signal with the small energy density of strings~(the decay products are still under debate~\cite{Vincent:1997cx,Hindmarsh:2008dw}, here we only discuss the Nambu-Goto strings). In this case, most GWs are produced from the decay of string loops which result from collisions of long strings. 
	The SGWB from string loops properly results in the nanohertz SGWB with $\mu/(8\pi)\sim 10^{-12}$~\cite{Ellis:2023tsl,Lazarides:2023ksx}, where $\delta_{H}\sim \mu$ is much lower than the threshold of subhalo formation.
	
	At the end of inflation, the condensate of inflaton oscillates in the minimum of the effective potential. Perturbations of inflaton are exponentially amplified by parametric resonance in the presence of strong interactions~\cite{Kofman:1997yn,Khlebnikov:1997di,Amin:2010dc,Zhou:2013tsa,Lozanov:2019ylm,Liu:2017hua}. 
	A similar scenario occurs for the light fields during inflation when the Hubble parameter becomes smaller than their effective mass~\cite{Olle:2019kbo,Inomata:2020xad,Kitajima:2020rpm,Ramberg:2020oct,Kawasaki:2021ycf}. Because of the nonlinear evolution, lattice simulations are required to obtain the GW energy spectra, which are approximately written in a unified form $\Omega_{\mathrm{GW}}^{\mathrm{cf}}\approx 0.1\Omega_{\mathrm{r}}(\frac{k_{\mathrm{res}}}{a_{*}H_{*}})^{-2}$~\cite{Lozanov:2019ylm}, where $k_{\mathrm{res}}$ denotes the wavenumber of energy density perturbations amplified by parametric resonance.
	Energy density perturbations reaches $\mathcal{O}(1)$ level at the scale of $k_{\mathrm{res}}$ and causality implies that $\delta\rho/\rho\propto k^{3/2}$ in the infrared region. Therefore, we obtain $\delta_{H}\sim  (\frac{k_{\mathrm{res}}}{a_{*}H_{*}})^{-\frac{3}{2}}$.
	
	Primordial curvature perturbations which originate from quantum fluctuations during inflation also generate GWs since scalar modes of metric perturbations are coupled to tensor modes at the second-order expansion of the Einstein equation. 
	Many non-slow-roll inflationary models predict large-amplitude primordial perturbations at small scales~(see Rev.~\cite{Ozsoy:2023ryl} for a recent review) so that scalar-induced GWs are strong enough to be observed by multiband GW detectors. The energy spectrum of scalar-induced GWs and $\delta_{H}$ depend on the specific form of the power spectrum of curvature perturbations, $P_{R}(k)$. Here, we simply apply the approximated expression, $\Omega^{\mathrm{pc}}_{\mathrm{GW}}\sim \Omega_{r}P_{R}^{2}$ and $\delta_{H}\sim\sqrt{P_{R}(k)}$. See Refs.~\cite{Press:1973iz,Kohri:2018awv,Espinosa:2018eve} for the explicit result. This issue has been considered before in, for example, Refs.~\cite{Delos:2023fpm,StenDelos:2022jld}, where the authors obtain constraints on $P_{R}$ from compact substructures of DM.
	\begin{figure*}
		\includegraphics[width=3.2in]{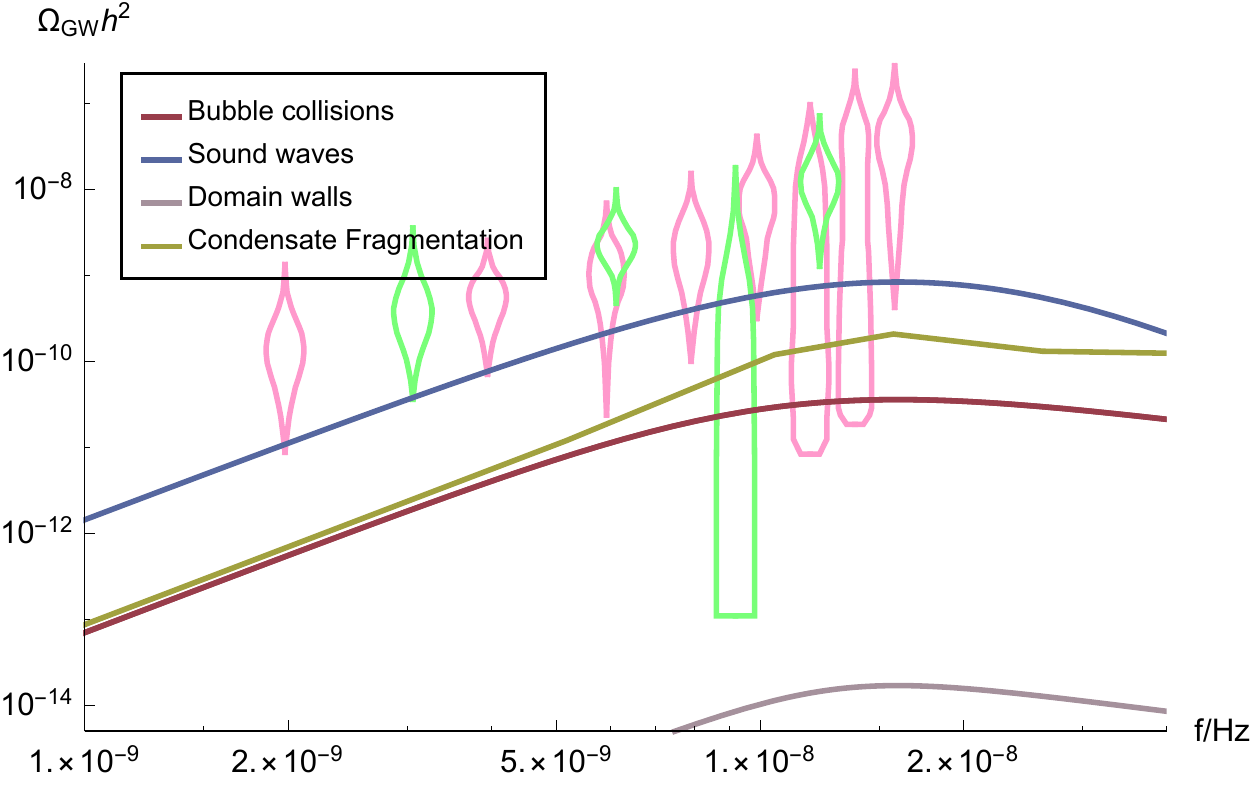}
		\includegraphics[width=3.3in]{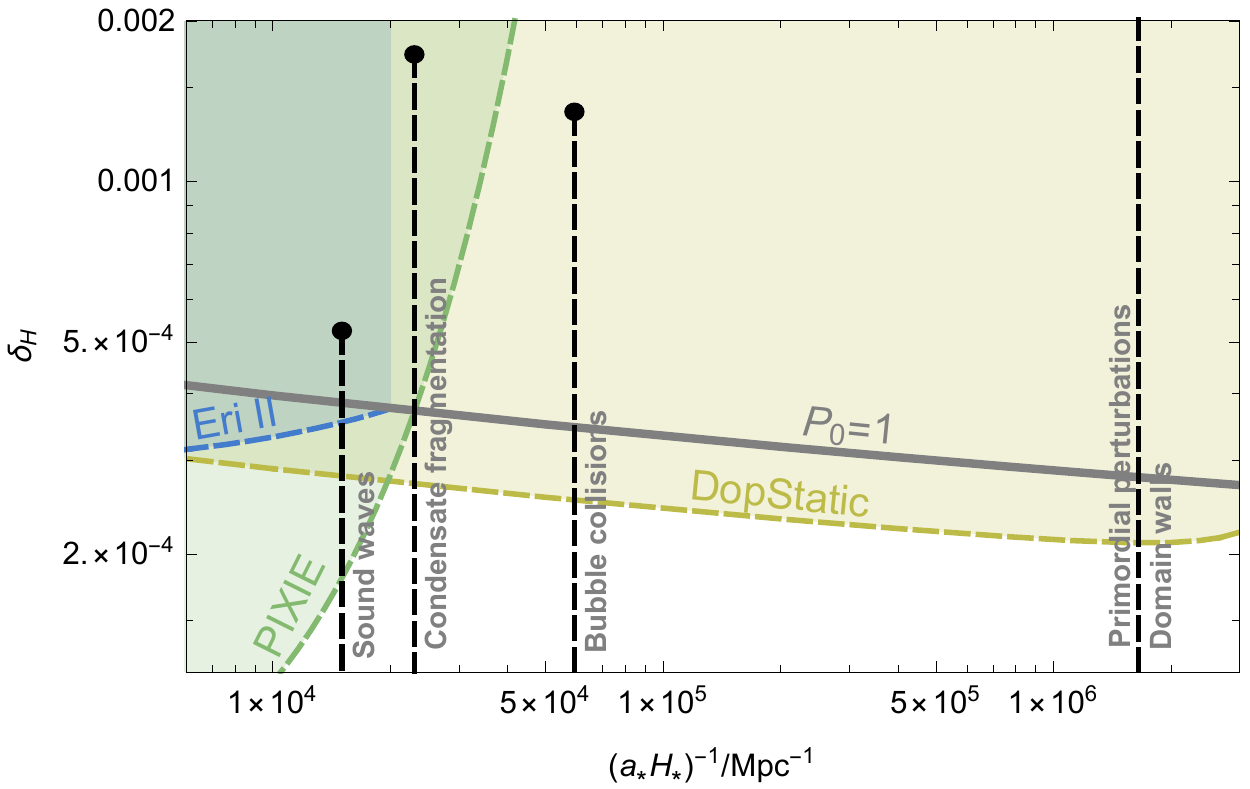}
		\caption{Left: The maximum $\Omega_{\mathrm{GW}}$ from each GW source~(the solid lines) assuming that $P_{0}$ is less than unity, which is compared to the constraints given by NANOGrav and EPTA~(the pink and green violin plots). Here, we do not show the case of primordial curvature perturbations since the GW signal is too weak. Right: The verticle dashed lines show the minima of induced $\delta_{H}$ when the GW sources in Tab.~\ref{tab:value} have the chance to explain the PTA signal, all of which exceed the threshold of $P_{0}=1$~(the solid gray line). The constraints are obtained from pulsar timing correlations~(dashed yellow line), CMB $\mu$-distortions~(dashed green line), and the population of ultra-faint dwarf galaxies~(dashed blue line).}
		\label{fig:gw}
	\end{figure*}
	
	\emph{The formation of DM subhalos}. 
	For the simplest collisionless cold DM model, DM collapses into compact subhalos at around matter-radiation equality if their energy density perturbations exceed a threshold of $\delta\sim 10^{-3}$~\cite{Berezinsky:2003vn,Berezinsky:2003vn,Berezinsky:2005py,Ricotti:2009bs}.
	The numerical result in Ref.~\cite{PhysRevD.97.041303} implies the DM subhalos have the Navarro-Frenk-White-like profiles for a spike-like $P_{R}(k)$
	\begin{equation}\label{eq:NFW}
		\rho(r)=\frac{\rho_{s}}{\left(r/r_{s}\right)^{3/2}\left(1+r/r_{s}\right)^{3/2}}\,.
	\end{equation}
	The parameters $\rho_{s}\approx30(1+z_{c})^{3}\rho_{\mathrm{DM},0}$, $r_{s}\approx(1+z_{c})^{-1}(a_{*}H_{*})^{-1}$ are fixed by simulations, where $\rho_{\mathrm{DM},0}$ is the averaged DM density at present and $z_{c}$ is the collapse redshift. The present halo mass $M_{0}$ is obtained by integrating $\rho(r)$ inside the viral radius $r_{\mathrm{vir}}$, where  $\rho(r_{\mathrm{vir}})=200\rho_{\mathrm{DM},0}$ by definition~\cite{Bryan:1997dn}. The compactness of the DM subhalos is described by the concentration parameter, $c=r_{\mathrm{vir}}/r_{s}$. Assuming that all of the subhalos form before $z_{c}=500$, we can obtain a conservative estimation $c=270$.
	
	Assuming Gaussian energy density perturbations, the present abundance of subhalos that formed before $z_{c}<500$ is then approximated as~\cite{Bringmann:2011ut}
	\begin{equation}\label{eq:P0}
		P_{0}=\min\left(\frac{M_{0}}{M_{i}}\frac{1}{\sqrt{2\pi}\delta_{H}}\int_{\delta_{min}}^{+\infty}d\delta\,\exp\left(\frac{-\delta^{2}}{2\delta_{H}^{2}}\right),1\right)\,,
	\end{equation}
	where $M_{i}=\frac{4}{3}\pi (a_{*}H_{*})^{-3}\rho_{DM,0}$ is approximately the total DM mass inside the Hubble horizon at the halo formation time. We follow Ref.~\cite{Bringmann:2011ut} to decide the threshold $\delta_{\mathrm{min}}$ with $z_{c}=500$. Note that, we do not require large isocurvature perturbations to produce abundant subhalos which depending on the DM generation mechanism~\cite{Hogan:1988mp,Kolb:1993zz,Arvanitaki:2019rax,Graham:2015rva,Erickcek:2011us}, but simply assume negligible isocurvature perturbations.
	
	
	
	\emph{Mutual constraints on GW sources and dark matter models}.
	We apply the constraints on $\Omega_{\mathrm{GW}}$ from the results of NANOGrav and EPTA, depicted as the pink and green violin plots respectively in the left panel of Fig.~\ref{fig:gw}.
	To simplify the data processing, we assume that the SGWB can only appropriately explain the observational data if $\Omega_{\mathrm{GW}}>10^{-9}$ at $16$ nHz.
	In the left panel of Fig.~\ref{fig:gw}, we show the maximum $\Omega_{\mathrm{GW}}$ throughout the parameter spaces of each GW source in the case $P_{0}<1$. The corresponding maximum values of $\Omega_{\mathrm{GW}}$ at $16$ nHz are listed in the second column of Tab.~\ref{tab:value}. 
	The result indicates that the GW sources in Tab.~\ref{tab:value} can only appropriately explain the PTA result with the near-unity abundance of subhalos that formed before $z_{c}=500$. Nambu-Goto string loops, unlike other GW sources, can generate strong GW signals while $\delta_{H}$ remains much less than $\delta_{\mathrm{min}}$ because of their high efficiency of GW production.
	
	We also obtain the lowest amplitude of $\delta_{H}$ for each GW source assuming they are strong enough to result in the nanohertz SGWB, as depicted in the right panel of Fig.~\ref{fig:gw}, all of which exceeds the threshold of $P_{0}=1$ shown in the solid gray line. The corresponding values of $M_{0}$ and $\delta_{H}$ are listed in the third and fourth columns of Tab.~\ref{tab:value}.
	For string loops, induced $\delta_{H}$ is very low which is not shown in Fig.~\ref{fig:gw}. 
	The observational constraints on the subhalo abundance from pulsar timing correlations~\cite{Ramani:2020hdo} and the population of ultra-faint dwarf galaxies~\cite{Brandt:2016aco}, have been converted to the upper limits on $\delta_{H}$ according to Eq.~\eqref{eq:P0}. 
	
	For the halo mass range considered in this work, the main contribution to the pulsar phase corrections comes from the static Doppler effect, where the pulsars accelerate due to the gravitational effect of static subhalos, affecting the second-order derivative of the pulsar frequency. The yellow dashed line in Fig.~\ref{fig:gw} shows the constraint from pulsar timing obtained in Ref.~\cite{Ramani:2020hdo} with $c=100$ and their optimistic parameters~(the number of pulsars $N_{P}=1000$, the root-mean-square post-fit timing residual $t_{\mathrm{rms}}=10$ ns, the measurement cadence $\Delta t=1$ week, the observation duration $T=30$ yr, the distance of the pulsars $z_{0}=10$ kpc), denoted as $DopStatic$. Although the optimistic condition is far future prospects, we only require a more realistic precision to observe the predicted mass function, such as $N_{P}=24$, $t_{\mathrm{rms}}=10$ ns, $\Delta t=$2 weeks, $T=$20 years, $z_{0}=5$ kpc. 
	
	Compact DM substructures can dynamically heat the star clusters near the core of the ultra-faint dwarf galaxies, and the dashed blue line shows the constraint from the survival of the star cluster in Eridanus II, denoted as $Eri\,II$~\cite{Brandt:2016aco}. The mean separation of the subhalos exceeds $20$ times $r_{s}$ even in the core of the dwarf galaxies, so the subhalos considered in this work can be safely treated as point-like objects.
	Note that the above constraints only rely on the gravitational effects of DM subhalos rather than specific interactions with the SM particles. For weakly interacting massive particles~(WIMPs), the compact subhalos enlarge the DM self-annihilation rate~\cite{Berezinsky:2003vn,PhysRevLett.103.211301}, so the non-detection of the corresponding gamma-ray signal from the Fermi satellite gives strict limitation on the halo abundance in the mass range $10^{-8}\,M_{\odot}\lesssim M_{0}\lesssim10^{12}\,M_{\odot}$~\cite{Bringmann:2011ut,PhysRevD.98.063527,FrancoAbellan:2023sby}, where $M_{\odot}$ denotes the solar mass. Then, if DM is mainly constituted from WIMPs and the final annihilation products are photons, all of the sources in Tab.~\ref{tab:value} are ruled out by this constraint alone.
	
	The observations of the CMB $\mu$-distortions directly limit induced curvature perturbations caused by GW sources, where $\mu$ is obtained in terms of $\delta_{H}$ following the method presented in~\cite{Chluba:2015bqa}. In this approach, we follow the Press-Schechter formalism~\cite{Press:1973iz} and employ a $\delta$-function $P_{R}(k)$ that peaks at $k=R_{H}^{-1}$ to derive a conservative constraint on $\delta_{H}$ . The Primordial Inflation Explorer~(PIXIE) has been designed with a sensitivity such that $\mu<3\times 10^{-8}$, which provides the constraint on $\delta_{H}$ illustrated by the dashed green line in Fig.~\ref{fig:gw}.
	
	In this work, we investigate the runaway scenario of the FOPTs in which $v_{w}\sim 1$. For $v_{w}<1$, GWs generated from sound waves and bubble collisions increase slightly with the same value of $\delta_{H}$. In such scenarios, we find that sound waves can appropriately explain the SGWB with $P_{0}<1$, while the conclusion remains unchanged for bubble collisions. Although observations of DM subhalos may not be able to eliminate sound waves, they could impose more stringent constraints on the model parameters.
	
	The observations of DM subhalos cannot impose strict constraints on cosmic string loops. Since $\Omega_{\mathrm{GW}}$ from string loops has a characteristic plateau, the cosmic string model parameters receive additional constraints on GW observers in other frequency bands, such as aLIGO and LISA~\cite{Hindmarsh:2022awe}~(For unstable strings or strings diluted by inflation, the constraint from LIGO-Virgo-KAGRA collaboration can be released~\cite{Ellis:2023tsl,Cui:2019kkd,Buchmuller:2020lbh}).
	\begin{table}[ht]
		\centering
		\begin{tabular}{|c|c|c|c|}
			\hline
			GW sources & $\Omega_{\mathrm{GW,m}}$&$M_{0}$&$\delta_{H}$\\
			\hline
			Bubble collisions & $3.6\times10^{-11}$&$0.331$&$1.4\times10^{-3}$\\
			Sound waves & $8.4\times10^{-10}$&$20.3$&$5.2\times10^{-4}$\\
			Condensate fragmentation & $1.4\times10^{-10}$&$5.62$&$1.7\times 10^{-3}$\\
			Domain walls & $1.7\times10^{-14}$&$ 1.5\times10^{-5}$&$2.1\times 10^{-2}$\\
			Primordial perturbations & $\sim 10^{-18}$&$1.5\times 10^{-5}$&$\sim 0.07$\\
			\hline
		\end{tabular}
		\caption{For each GW source in this table, $\Omega_{\mathrm{GW,m}}$ denotes the maximum value of $\Omega_{\mathrm{GW}}$ evaluated at $16$ nHz under the condition $P_{0}<1$, $M_{0}$ denotes the present halo mass in terms of the solar mass when the GW sources can result in the SGWB while inducing the lowest density perturbations at the Hubble horizon scale, $\delta_{H}$.}
		\label{tab:value}
	\end{table}
	
	With the help of future improved PTA sensitivity and the development of multiband GW observers, it is possible to differentiate the GW sources by precisely measuring the profile of $\Omega_{\mathrm{GW}}$ alone. Then, if the cosmological sources in Tab.~\ref{tab:value} have been verified, one would expect the existence of very abundant DM subhalos. In this case, the observation of nanohertz GWs offers a novel way of testing the nature of DM. WIMPs that decay into photons are strongly restricted. 
	The existence/absence of these subhalos in future observations would exclude/verify the DM models that prevent halo formation at small scales, such as warm DM and fuzzy DM models.
	The core radius of subhalos depends on the mass and annihilation rate of DM, allowing the measurement of the halo profile to detect DM model parameters without relying on specific interaction terms. 
	
	\emph{Conclusion and discussion}.
	We investigate the DM subhalo formation which results from energy density perturbations associated with the cosmological GW sources.  For cosmological GW sources including bubble collisions and sound waves during the FOPTs~(in the runaway scenario), domain walls, condensate fragmentation, and primordial curvature perturbations from inflation, if GWs from these sources are strong enough to explain the PTA data, induced energy density perturbations will lead to the formation of subhalos with near-unity abundance. 
	Since the current observational error is too large to determine the source from $\Omega_{\mathrm{GW}}$, the observations of subhalos provide a novel method to distinguish the GW sources and give additional constraints on the model parameters.
	Conversely, if future PTA experiments confirm one of these GW sources from the GW energy spectrum, the collisionless cold DM model predicts the existence of very abundant subhalos. In this case, we can constrain the DM model parameters, such as the temperature, the mass, and the interactions, from the observations of the halo abundance and profile instead.
	
	The result of this work is not restricted to the nanohertz GW sources.
	The authors of Ref.~\cite{Ramani:2020hdo} find the observations of pulsar timing corrections have the potential to constrain the abundance of subhalos in a large mass range with $M_{0}\gtrsim 10^{-13}\,M_{\odot}$, which has the ability to constrain the GW source in the $\mu$Hz frequency gap where the current methods can hardly detect~(see Refs.~\cite{Sesana:2019vho,Blas:2021mqw,Bustamante-Rosell:2021daj} for new methods). On the other hand, the upper bounds on $\Omega_{\mathrm{GW}}$ given by laser interferometer GW detectors, such as aLIGO and LISA, also places upper bounds on $P_{0}$ for $10^{-4}\, \mathrm{g}\lesssim M_{0}\lesssim 10^{20}\, \mathrm{g}$.
	The observations of GWs and curvature perturbations together yield a more general joined measurement of the early Universe.
	
	\emph{Acknowledgments}
	We sincerely thank Zong-Kuan Guo and Ligong Bian for the fruitful discussions.
	This work is supported in part by the National Key Research and Development Program of China Grants No. 2020YFC2201501 and No. 2021YFC2203002, in part by the National Natural Science Foundation of China Grants No. 12105060, No. 12147103, No. 12235019, No. 12075297 and No. 12147103, in part by the Science Research Grants from the China Manned Space Project with NO. CMS-CSST-2021-B01,
	in part by the Fundamental Research Funds for the Central Universities. 
	
	
	
	
	\bibliography{PTAPR}
\end{document}